\documentclass[a4paper,12pt]{article}
\usepackage[utf8]{inputenc}
\usepackage{cancel}
\usepackage{ulem}
\usepackage{amsfonts}
\usepackage{amssymb}
\usepackage{graphicx}
\usepackage{amsmath}
\usepackage{enumerate}
\usepackage{here}
\usepackage{subfloat}
\usepackage{subfig}
\usepackage{color}
\usepackage{float}
\usepackage{here}
\usepackage{cite}
\usepackage{mathrsfs}
\usepackage{float,epsfig}
\usepackage{dcolumn}
\usepackage{multirow}
\usepackage{placeins}
\usepackage{graphicx}
\usepackage{bm}
\usepackage{amsmath,amssymb,amsthm}
\usepackage[colorlinks=true,linkcolor=blue]{hyperref}
\hypersetup{citecolor=magenta}
\usepackage[table]{xcolor}
\title{\bf \Large{
Particle emission spectrum and thermodynamics phase transition of  RN-AdS black hole in massive gravity  via a new prescription 
}}

\author{Mohamed Chabab$^{1,2}$\footnote{mchabab@uca.ac.ma } ,
\  Samir Iraoui$^{1,3}$\footnote{samir.iraoui@ced.uca.ma},
\ Hicham Sriba$^1$\footnote{h.sriba.ced@uca.ac.ma}\\
	\\ 
	{\small $^1$ Cadi Ayyad University, FSSM,  High Energy and Astrophysics Laboratory,
	}\\
	{\small  P.O.B. 2390,  Marrakech 40000, Morocco.
	}\\
{\small  $^2$ Cadi Ayyad University, National School of Applied Sciences,   P.O.B. 63 , } \\
{ \small  Safi 46000, Morocco} \\ 
{\small  $^3$ Centre Regional des Metiers de l'Education et de la Formation de Marrakech,}\\
{ \small  Marrakech 40000, Morocco} 
} 

\date{}

\begin{document} 
\maketitle	
\begin{abstract}
	In this paper we investigate the shadow and energy emission rate of a charged AdS black hole in massive gravity. We develop a new prescription based on maximum frequencies of emission spectrum to probe the thermodynamics of  black holes and get an accurate insight into its phase transition criticality. We establish a link between the thermodynamic behavior and the maximum emission frequency of the black hole for both the photons and massive bosonic modes.  We show that the  frequency maximizing the spectrum combined to the shadow radius can serve as a powerful physical tool to delve into the thermodynamics and phase structure of RN-AdS black holes in massive gravity.

\end{abstract}


          \tableofcontents
  
\renewcommand{\thefootnote}{\arabic{footnote}}
\setcounter{footnote}{0}
\section{Introduction}
\label{intro}

The Event Horizon Telescope Collaboration released the first image of M87 \cite{EHT:2019nmr,EventHorizonTelescope:2019ths,EventHorizonTelescope:2019dse} in 2020. This is the first-ever image of the shadow of a black hole, from which the mass and speed of rotation of the black hole were calculated. Since then, the study of black hole shadow received an increasing interest  and has become an  active research topic. If the objects moving around black holes are photons coming from an illuminated source, it casts a shadow in a plane which can be seen by an observer at infinity. The black hole shadow was first studied by Bardeen \cite{Bardeen:1973tla}. Then, it was widely analysed in many gravitational models where a plethora of papers with interesting results on the shape of  black hole shadows  were reported \cite{Luminet:1979nyg,Amarilla:2010zq,Afrin:2021imp}.\\
		
The study of  the shadow proved to be useful to uncover information encoded in a black hole and some of its features by observing the behavior of matter in the vicinity of its outer horizon. Recently,  a large number of papers dealing with the geodesics of a particle around a black hole within various theories of gravity~\cite{Pugliese:2010ps,Wei:2017mwc}. So contributing to deepen our  understanding of many observational effects such as the gravitational deflection angle of light and time delays~\cite{Fu:2022yrs,Ezquiaga:2020dao}.\\
	
Hawking has pointed out that black holes can emit particles in thermal form, alike the  black body radiation. This notably originates from the quantum tunneling created by vacuum fluctuations near the black hole's event horizon. Therefore,  when quantum effects  and the laws of thermodynamics are considered, the  black holes can produce radiation, a phenomenon dubbed Hawking radiation \cite{Hawking:1974rv,Hawking:1982dh,Hawking:1975vcx
}.  Page~\cite{Page:1976df} combines Hawking's quantum formalism with the  perturbation theory to calculate, for the first time, the emission rate of massless particles. In this context, Emparan et al.\cite{Emparan} showed  that most of the energy is radiated into the Standard Model particles on the brane rather than Kaluza-Klein modes on the bulk in extra dimensions spacetime.  This could pave the way to observing Hawking radiation and measuring  the emission rate in the future high energy colliders. \\

Thanks to the pioneering work of Hawking, Carter, Bardeen and Page in the seventies~\cite{Bardeen:1972fi,Bardeen:1973tla,Bardeen:1973gs}, the  black holes thermodynamics emerge as a rich fundamental research field. More particularly, their extended phase space has attracted increasing attention in recent decades. The key idea comes from identifying the cosmological constant as thermodynamic pressure. This implies a phase structure analogous to that of Van der Waals \cite{Kastor:2009wy,Dolan:2011xt,Cai:2013qga,Kubiznak:2012wp}. In this context, various techniques were proposed to consolidate this analogy, such as the thermodynamic geometry \cite{Weinhold:1975xej,Ruppeiner:1995zz,Quevedo:2006xk} and quasi-normal modes \cite{Chabab:2016cem,Chabab:2017knz,Mahapatra:2016dae}. \\

Moreover, many papers have also focused on the generalization to alternative theories of gravity,  especially in the massive gravity \cite{deRham,deRhamdRGT,HassanF,Ghosh}. More recently, many efforts have been made to correlate thermodynamic behavior of a black holes with its shadow \cite{Shaikh:2018lcc,Wang:2021vbn,EslamPanah:2020hoj}. The aim of the present work is to study the shadow and the energy emission rate of charged AdS black holes in the massive gravity background. To this end, we propose a new prescription which rely especially on an observable quantity, namely the maximum emission frequency, rather than on the usual geometric quantities as the event radius  the black hole.\\
		
This paper is organized as follows. In the next section, we present a concise review of the thermodynamic   of  massive gravity charged black holes in AdS space. In section~\ref{sec3},  we first establish the equation of motion for a free photon propagating in the equatorial plane around RN-AdS black hole in massive gravity. Then, we examine its shadow.  In section~\ref{sec4}, the emission rate of energy is analysed for either a massless or a massive field, with an emphasis on its connection with the thermodynamical phase transition of black holes.  Finally, the last section is devoted to highlight our results.

  \section{Thermodynamics of charged AdS black holes in massive gravity}\label{sec2}
  
We study solutions of $4$D charged-AdS black hole in massive gravity theory described by the following action \cite{Chabab:2019kfs,Zeng:2015tfj,Hendi:2015eca}
 \begin{equation}
     S= \int\, \mathrm{d}^4x\sqrt{-g}\left[ \mathcal{R} - 2 \Lambda - \frac{1}{4}\mathcal{F}^2 +m_g^2\sum_{i = 1}^{4}c_i\mathcal{U}_i(g,f)\right] ,
 \end{equation}
where $\mathcal{F}= F_{\mu\nu}F^{\mu\nu}$ is the Maxwell invariant and $F_{\mu\nu}$ is  electromagnetic tensor, $\mathcal{R}$ is the Ricci scalar, $m$ is the graviton mass and $\mathcal{U}_i$ stand for symmetric polynomials of the eigenvalues of  $4\times4$ matrix defined as
\begin{equation}
    \mathcal{K}_{\nu}^{\mu}=\sqrt{g^{\mu\alpha}f_{\alpha\nu}}.
\end{equation}

Following Cai et al. \cite{Cai:2021fpr,Hendi:2015hoa} we can choose the reference metric as follows
\begin{equation}
f_{\mu\nu} = diag\left(0,0,c_0^2h_{ij}\right),
\end{equation}    
leading to the charged AdS black hole solution in massive gravity
  \begin{equation}
     ds^2 = -f(r)dt^2 + \frac{dr^2}{f(r)} + r^2 \left(d\theta^2 + r^2 \sin^2{\theta}d\phi^2\right).
 \end{equation}
 The metric function has the form
\begin{equation}
    f(r) = 1 - \frac{2M}{r} + \frac{q^2}{4r^2} + \frac{r^2}{l^2} + m_g^2( r\frac{c_0c_1}{2} +  c_0^2 c_2),
    \label{eq5}
\end{equation}
where $M$, $q$ represent the mass and electric charge of the black hole respectively, while $l$ is the  AdS length scale set by the presence of the cosmological  constant $\Lambda$. Subsequently, it would be useful to use the following notations: $a=m_g^2\frac{c_0c_1}{2}$ and $b=m_g^2c_0^2c_2$.

The cosmological constant mimics a thermodynamic variable identified to the pressure of a black hole \cite{Kastor:2009wy} by the relation,
\begin{equation}
P= \frac{3}{8\pi l^2}.
    \label{eq}
\end{equation}
The mass of the black hole  reads
\begin{equation}
    M=\frac{3\pi^{3/2} q^2 + 12a S^{3/2} + 4\sqrt{\pi} S(3+3b+8 P S)}{24\pi\sqrt{S}},
\end{equation} 
where $S$ is  the entropy of the black hole, defined by $S=A_H/4$ where $A_H$ is the horizon area. The conjugate quantities of these intensive parameters can be readily deduced from the following equations~\cite{Chabab:2019kfs,Hendi:2017bys,Kubiznak:2012wp,Chabab:2019mlu},
\begin{equation}
V=\frac{\partial M}{\partial P}|_{S,q,a,b}\quad ;  \quad \Phi=\frac{\partial M}{\partial q}|_{S,P,a,b} \quad ;  \quad \mathcal{A}=\frac{\partial M}{\partial a}|_{S,P,q,b}
\quad ;  \quad \mathcal{B}=\frac{\partial M}{\partial b}|_{S,P,q,a}.
\end{equation}
The Hawking temperature  related to  the surface gravity is given by
\begin{equation}
    T_H=\frac{-\pi^{3/2} q^2 + 8aS^{3/2} + 4\sqrt{\pi} S(1+b+8PS)}{16\pi S^{3/2}}.
    \label{eq9}
\end{equation}
Using Eq. \ref{eq9},  one can derive the pressure as a function of the temperature $T$ and the horizon radius $r_h$,
\begin{equation}\label{stateP}
    P(r_h)=\frac{q^2 - 4(1+b)r_h^2 - 8(a-2\pi T)r_h^3}{32 \pi r_h^4}.
\end{equation}

In the context of massive gravity, the generalized first law of thermodynamics can then be expressed as, 
\begin{equation}
    dM=TdS+\Phi dq + VdP + \mathcal{A}da + \mathcal{B}db,
\end{equation}
  while the Smarr relation is readily obtained by means of the  Euler's theorem \cite{Smarr:1972kt}:
\begin{equation}
    M= 2TS - 2VP + \Phi q - \mathcal{A}a.
\end{equation}

Next we focus on  the thermodynamical criticality of RN-AdS BH in massive gravity background. To this end, using the equation of state, we first determine the critical points: 
\begin{equation}
    T_c=\frac{a}{2\pi} + \frac{2(b+1)^{3/2}}{3\pi\sqrt{6}q}   ; \qquad
    P_c=\frac{(b+1)^2}{24\pi q^2}     ;\qquad
    r_c=\frac{\sqrt{\frac{3}{2} q}}{\sqrt{1+b}}.
\end{equation}
Then, we calculate the Gibbs free energy $G=M-TS$,
\begin{equation}
	G=\frac{9\pi q^2 + 4S(3+3b-8PS)}{48\sqrt{\pi}\sqrt{S}}.
\end{equation}

\begin{figure}[h]
  \begin{center}
    \includegraphics[scale=0.7]{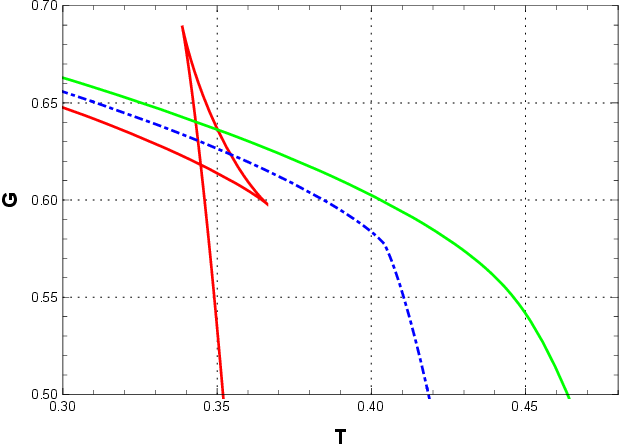}  
    \includegraphics[scale=0.49]{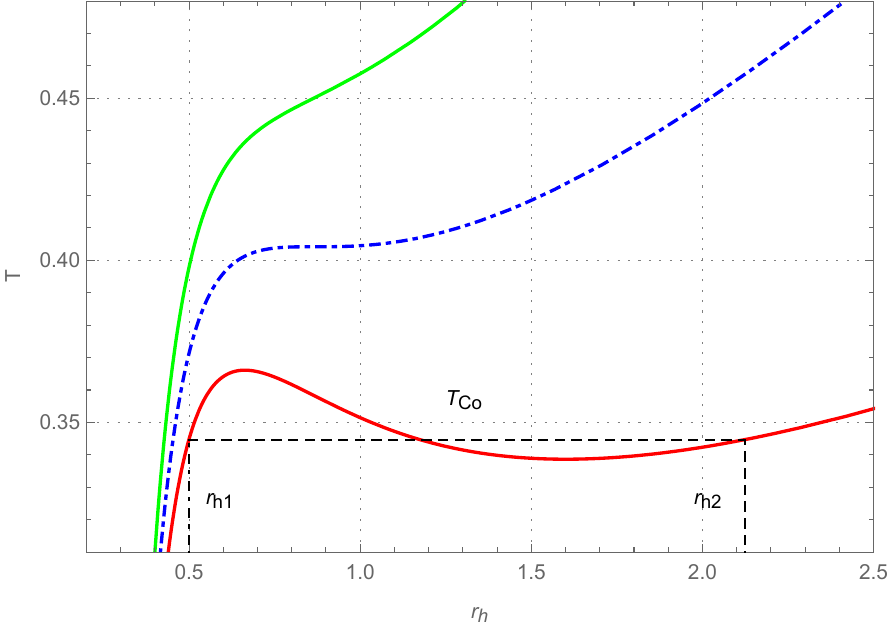}
  \end{center}
   \caption{\textbf{Left:} Gibbs free energy as a function of  the Hawking temperature $T$. \textbf{Right:} Thermodynamic behavior in $T-r_h$ diagram. The red curves correspond to  $P < P_c$, the blue curves to $P > P_c$ and black dashed is for $P=P_c$. We set $a=1,\ b=1$  and $q=1$.}
  \label{GT}
\end{figure}

To investigate the possible thermodynamical phase transitions, we have evaluated and plotted different thermodynamical quantities.  Fig.~\ref{GT} illustrates  the variations of Gibbs free energy as a function of temperature, and the pressure versus the horizon radius. We display, in the left panel of Fig.~\ref{GT}, $(G-T)$ diagram for different values of pressure. We can see that when $P$ is below  the critical pressure $(P < P_c)$, Gibbs free energy shows a swallowtail structure indicating occurrence of a first order phase transition. However,  if $P > P_c$, this behavior vanishes and  no phase transition can occur.

 Fig.~\ref{GT} shows the $T-r_h$ diagram where at the critical point the curve presents an inflexion point  separating  the supercritical and subcritical phase.
 For $P>P_c$, the temperature indicates a monotonic behavior, which shows that no phase transition is possible. For $P=P_c$ the black hole is in a critical state, below which there are two local extremes, thus a non-monotonic behavior. The region where $r_h < r_{h_1}$  ($r_h>r_{h_2}$) corresponds to the small (large) stable black hole, while the part where $r_{h_1}<r_h<r_{h_2}$ is the unstable phase. $T_{Co}$ is the coexistence temperature where the two different small/large BH phases can exist in equilibrium. $T_{Co}$ is calculated by equalizing the Gibbs free energies of the two phases~\cite{Chabab:2019kfs,Kubiznak:2012wp}.
 
\section{Shadow and thermodynamics of RN-AdS charged black holes in massive gravity}
\label{sec3}
This section is devoted to studying the shadow of  black hole with emphasis on its correlation with its thermodynamics.  First, we consider  a free massless particle, here a photon, whose motion and dynamics are described within the Lagrangian formalism,
\begin{equation}\label{largangiian}
\mathcal{L}=\frac{1}{2}g_{\mu\nu}\frac{\mathrm{d} x^{\mu} }{\mathrm{d} \sigma}\frac{\mathrm{d} x^{\nu} }{\mathrm{d} \sigma},
\end{equation}
with the coordinate $x^{\mu}=\left(t,r,\theta,\phi\right)$ and  $\sigma$ an affine parameters. The 4-momentum  reads as:
\begin{equation}\label{momentum}
p_{\mu}=\frac{\partial \mathcal{L}}{\partial \dot{x^{\mu}}}.
\end{equation}
Here, the dot denotes a derivative with respect to $\sigma$. Without loss of generality, we will restrict our analysis to the equatorial geodesics by setting  $\theta=\frac{\pi}{2}$ and  $\dot{\theta}=0$.  The spacetime symmetries have two Killing vectors $\partial_t$ and  $\partial_\phi$, implying  the existence of two constants of motion.
\begin{equation}
p_{t}=-E, \qquad p_{\phi}=L,
\end{equation}
where $E$ and $L$ represent the energy and $z$-component  of the angular momentum, respectively \cite{Li:2023zfl}.
The Euler-Lagrange equations with respect to $t$ and $\phi$ are readily obtained as
\begin{equation}\label{motioneuqationpt}
\frac{\mathrm{d} t}{\mathrm{d} \sigma}=\frac{E}{f(r)},
\end{equation}
\begin{equation}\label{motioneuqationpphi}
\frac{\mathrm{d} \phi}{\mathrm{d} \sigma}=\frac{L}{r^{2}}.
\end{equation}

Next, we introduce the effective potential by recalling the normalization condition for the null geodesics $g_{\mu\nu}\dot{x^{\mu}}\dot{x^{\nu}}=0$.  By considering this constraint,  the radial motion simplifies to the following simple form
\begin{equation}\label{Hamiltonien}
\dot{r}^{2}+V_{eff}(r)=0,
\end{equation}
where the effective potential $V_{eff}$  is given by
\begin{equation}
V_{eff}(r)=f(r)\left(\frac{L^2}{r^2} - \frac{E^2}{f(r)}\right).
\end{equation}
From Eq.~\eqref{motioneuqationpphi} and Eq.~\eqref{Hamiltonien},  the orbit equation for the photon reads as
\begin{equation}
\frac{dr}{d\phi}=\pm r \sqrt{f(r)\left(\frac{r^2 E^2}{L^2f(r)} - 1\right)}.
\label{equ25}
\end{equation}

The boundary of the shadow is mainly determined by the circular photon orbit satisfying the condition
\begin{equation}
V_{eff}(r_p)=\left.\frac{\partial V_{eff}}{\partial r}\right|_{r_p} = 0.
\label{equ23}
\end{equation}
While the turning point of the photon orbit can be derived via the  equation.
\begin{equation}
\left.\frac{dr}{d\phi}\right|_{r=r_p}=0,
\end{equation}
which yields  the relation
\begin{equation}
 \frac{dr}{d\phi}=\pm r \sqrt{f(r)\left(\frac{r^2 f(r_p)}{r_p^2f(r)} - 1\right)}.
    \label{equ27}
\end{equation}

 Consider the black hole shadow detected by the observer located at $r_0$ with an angular $\alpha$ relative to the radial direction,  one gets:
 
\begin{equation}
\cot\alpha=\left.\frac{\sqrt{f_{rr}}}{\sqrt{f_{\phi\phi}}} \frac{dr}{d\phi}\right|_{r=r_0}.
    \label{equ28}
\end{equation}
Then using Eq.~\eqref{equ27} we also obtain,
\begin{equation}
\sin^2\alpha=\frac{f(r_0)r_p^2}{r_0 f(r_p)}.
    \label{equ30}
\end{equation}
We see that the angular radius of the black hole shadow is readily evaluated as $r$ tends to a circular radius of the photon. Thus the shadow seen by a static observer at position $r_0$ is simply expressed by,
\begin{equation}
r_s=r_0 \sin\alpha=r\left.\sqrt{\frac{f(r_0)}{f(r)}}\right|_{r\to r_p}.
    \label{31}
\end{equation}
According to \cite{Shaikh:2019fpu,Xu:2018mkl,Hamil:2023dmx} apparent shape of the black hole shadow in the perpendicular plane of the observers direction, where $\theta_0=\pi/2$, is obtained by the celestial coordinates $x$ and $y$. The latter are generally defined for an observer far away from the black hole as:
\begin{align}
    x &= \lim_{r_0 \to \infty} \left. -r_0^2 \sin{\theta_0}\frac{d\phi}{dr}\right|_{(r_0,\theta_0)}, \\
      y &= \lim_{r_0 \to \infty} \left. r_0^2\frac{d\theta}{dr}\right|_{(r_0,\theta_0)}.
    \label{equ32}
\end{align}
\begin{figure}[h]
  \begin{center}
     \includegraphics[scale=0.8]{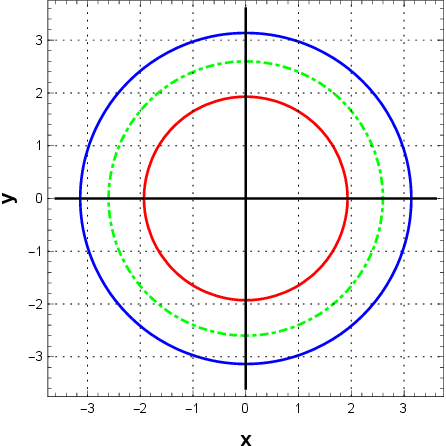}
  \end{center}
   \caption{Shape of the shadow cast by RN-AdS black hole in massive gravity for different pressures: $P<P_c$ (red), $P>P_c$ (blue) and $P=P_c$ (green dashed). Here we set $M=1$, $q=1$, $r_0=20$, $a=1$ and $b=1$. }
  \label{shadow}
\end{figure}

Fig.~\ref{shadow} illustrates how the black hole shadow behaves depending on the pressure value. One can see two main features: $1)$ The geometry of shadows remains  perfectly circular due to the absence of rotating parameter. $2)$ The pressure of the black hole increases with increasing of its shadow radius. 

Hereafter we focus on observable related to shadow that would  reflect the phase structure of charged AdS-black hole in massive gravity.  The relation between the black hole mass  and the size of its shadow can be written as
\begin{equation}
    r_s=A \sqrt{\frac{12A^2 f(r_0)}{32p\pi A^4 + 12A^3 + 24 A^2-24MA +3}}\ ,
    \label{rs}
\end{equation}
 where $A$ is a long-term function of $M$. 

\begin{figure}[h]
  \begin{center}
    \includegraphics[scale=0.695]{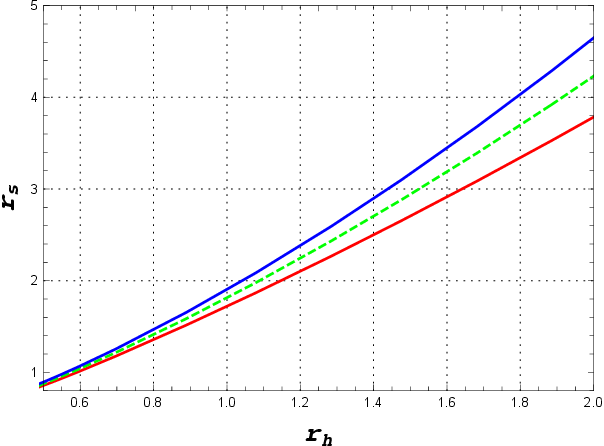} \includegraphics[scale=0.5]{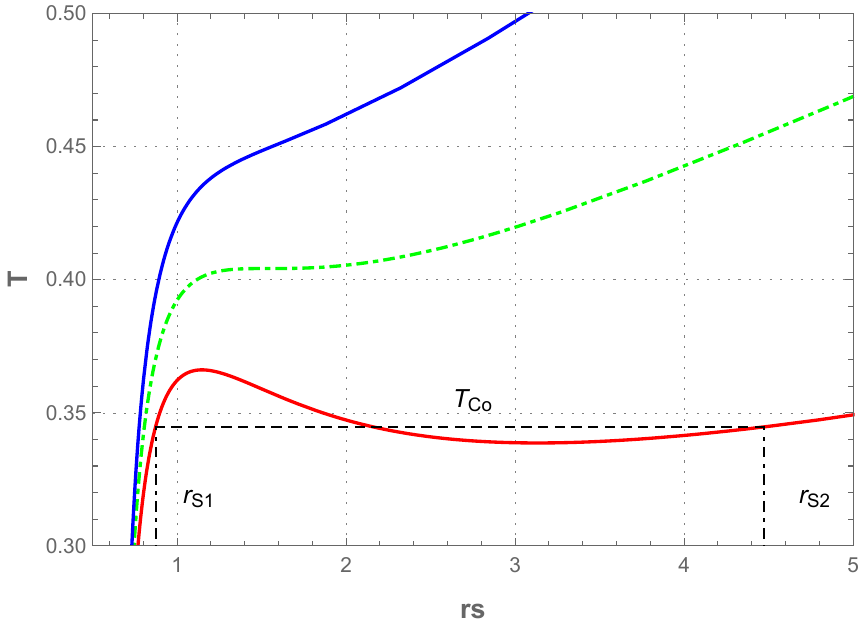}
  \end{center}
   \caption{\textbf{Left:} Shadow radius  as a function of the horizon. \textbf{Right:} Plot of temperature  versus the shadow radius. Red curves are below the critical point, the blue ones above the critical point and the green dashed are at the critical point. Here the parameters are  fixed as: $q=1$, $a=1$, $b=1$. The observer is in the equatorial plane and at position $r_0= 20$.}
  \label{rsrh}
\end{figure}

 The left panel of Fig.~\ref{rsrh} shows the black hole shadow radius $r_s$ as a function of  its horizon radius  $r_h$. We see that $r_s$ and $r_h$ display a quasi correlation, which is independent of the pressure. So  we can substitute the horizon radius by the shadow radius  in Eq.~\eqref{eq9} expressing the temperature of black hole. The quasi correlation between $r_h$ and $r_s$ suggests that the shadow radius can also reveal the phase structure of the spherically symmetric black hole. Indeed, the sign of quantity $\partial T/\partial r_h$ is equal to that of $\partial T/\partial r_s$ as reported  in~\cite{Zhang:2019glo}.

In the right panel of  Fig.~\ref{rsrh}, we display the temperature as a function of the shadow radius for different values of the pressure around $P_c$.  We can see a behavior that mimics the structure illustrated in the plane $T-r_h$: Similar phase transition is reproduced using the shadows of black hole, where phase transition and unstable region appearing in the interval $r_{s_1}<r_s<r_{s_2}$.\\

Next we explore the thermodynamical stability of the RN-AdS BH in massive gravity by means of  the heat capacity $C_P$.  The latter is evaluated via the  formula,
\begin{equation}
    C_P=T\left(\frac{dS}{dT}\right)_P 
    =\frac{-2 \pi q^2 r_h^2 + 8 \pi r_h^4 \left[1 + b + 2 r_h \left(a + 4 p \pi r_h\right)\right]}{3 q^2 - 4 r_h^2 (1 + b - 8 p \pi r_h^2)}.
\label{cp}
\end{equation}
The sign of heat capacity $C_P$ can uncover the stability structure, while its divergencies provide the phase transition critical points \cite{Davies}.

\begin{figure}[h]
	\begin{center}
		\includegraphics[scale=0.9]{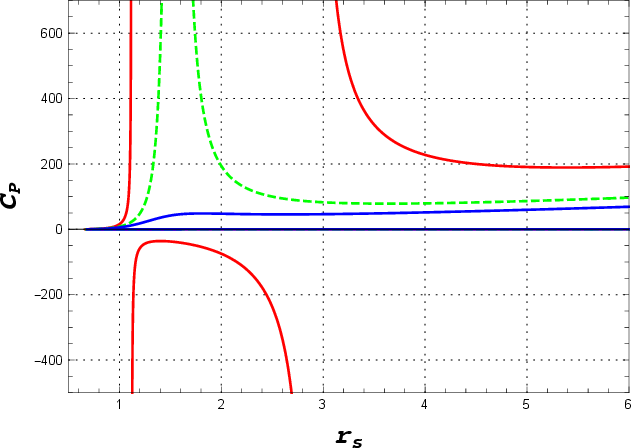}
	\end{center}
	\caption{Heat capacity as a function of the shadow radius $r_s$. The observer is at a distance $r_0= 20$ from the black hole. Color coding is similar to previous figures. Here we used $q=1$, $a=1$, $b=1$. }
	\label{cprs}
\end{figure}

In light of our new prescription, we analyze the heat capacity in the plane ($C_P-r_s$), with the shadow radius instead of the horizon radius. Again we recover  the same trend and results as in the $C_P-r_h$ plane, which consolidate further our approach based on observables related to black hole shadow as an efficient means to delve into the intricacies of its black holes thermodynamics. This study is shown in  Fig.~\ref{cprs}.

\section{Thermodynamics  and the  emission fields}\label{sec4}
Here we study  the emission rate of a massive field in four dimensions with emphasis on the frequency behaviour. Our objective is to establish a connexion with the thermodynamic behaviour of charged AdS black holes in massive gravity. Since, the maximum emission frequencies and the shadow radius are physical observables, we also aim to establish a new approach to probe the thermodynamics of black holes in massive gravity. In addition, as a by-product, we examine how the Wein's displacement law for the black holes evolves in terms of frequency, with either a massless particle or massive graviton. 

The energy emission rate at approaches of black hole shadow. Generally,
for a far distant observer, the high energy absorption cross section oscillates around a limiting constant value  approximated as:
 \begin{equation}
\sigma_{\text{lim}}=\pi r_s^2,
\label{equ36}
 \end{equation}
where $r_s$ is the shadow radius of the black hole.

\begin{figure}[h]
	\begin{center}
		\includegraphics[scale=0.69]{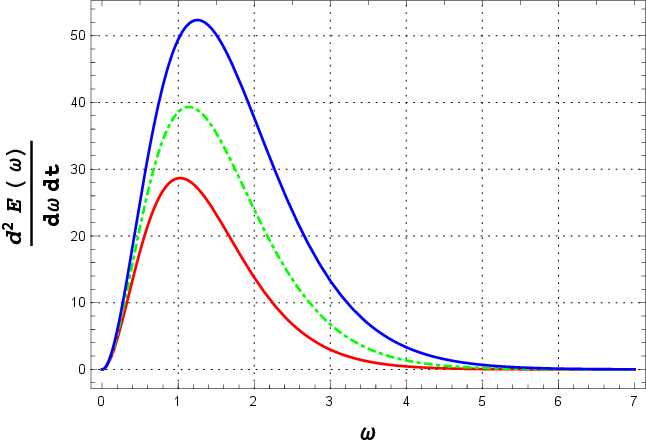} 
		\includegraphics[scale=0.69]{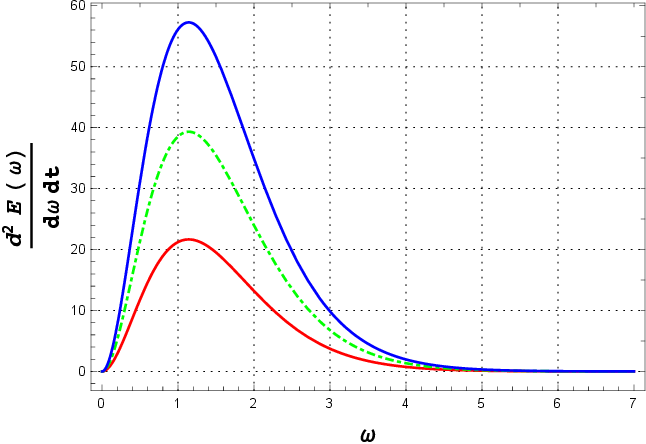}
		\caption{The emission spectrum of photons for different values of pressure and temperature. \textbf{Left:} For $P$ fixed, red curve is for $T< T_c$  the green dashed is for $T=T_c$, and the blue curve for $T>T_c$. \textbf{Right:} For $T$ fixed,  red curve for $P< P_c$, green dashed is for $P=P_c$, and blue curve for $P>P_c$. Here we set $a=1$, $b=1$, $q=1$, $r_0=20$ and $M=1$.}
		\label{Ew}
	\end{center}
\end{figure}

The  energy emission rate is generally expressed by the formula\cite{Ovgun:2019jdo,EslamPanah:2020hoj}
\begin{equation}
    \frac{d^2E(\omega,m)}{d\omega dt}= \frac{2\pi \sigma_{\text{lim}}}{e^{\omega/T_H} -1} \omega^3,
\end{equation}
where $T_H$ is the Hawking temperature  and $\omega$ is the frequency of the emitted photon. Fig.~\ref{Ew} illustrates the  emission spectrum for different values of temperature and pressure.  As one can see, the maximum of energy emission rate gets larger when the  pressure and temperature increase.  In addition, we also observe from the left panel of Fig.~\ref{Ew} that it shifts toward low frequencies direction if the temperature is augmented. From the right panel, we see that the frequency corresponding to the maximum emission $\omega_{max}$ does not change with pressure. This suggests that $ \omega_{max}$  depends essentially on the temperature. 
 
 The maximum of the spectral distribution in terms of the temperature is evaluated from the equation,
 \begin{equation}
 \frac{d}{d\omega}(\frac{d^2E(\omega)}{d\omega dt}) = 0.
 \label{omga max}
 \end{equation}
 Consequently, we can duly derive the Wien's displacement law, $ \omega_{max}=2,82\, T_H$\footnote{It is worth noticing that the generalized Wien's displacement law in $D+1$ spacetime dimensions  reads as: $\omega_{max}/T_H=D+W\left(-De^{-D}\right) $, where $W$ is the Lambert function\cite{Cardoso:2005cd}.}. As a result, the maximum frequencies of the emitted photons  can play the same role as the temperature in revealing the thermodynamics and criticality of black holes.

\begin{figure}[h]
	\begin{center}
		\includegraphics[width=.48\linewidth]{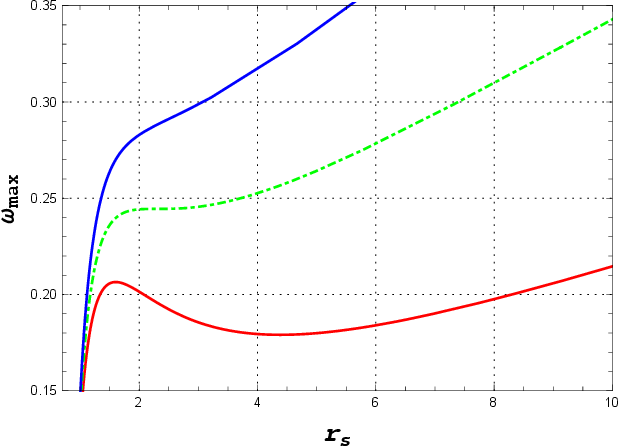}
		\includegraphics[width=.48\linewidth]{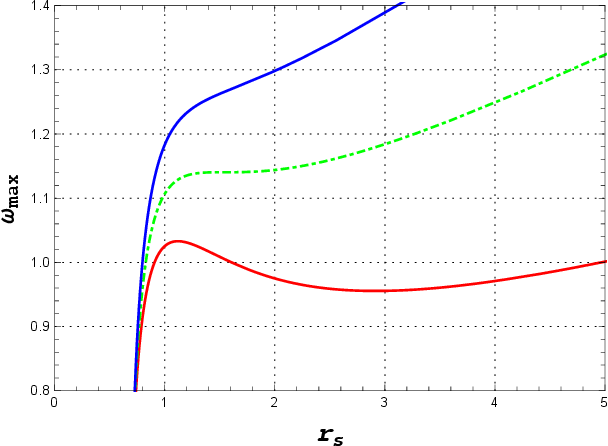}
		\caption{Maximum frequency of the  energy emission rate as a function of  the shadow radius for isobaric process. \textbf{Left:} RN-AdS black hole without massive gravity. \textbf{Right:} RN-AdS black hole in massive gravity with the parameters  fixed to the values $a=1$ and $b=1$. here we also used $q=1$ and $r_0=20$.}
		\label{wmaxrs}
	\end{center}
\end{figure}

In Fig.~\ref{wmaxrs}, we  plot the maximum frequency of the  energy emission rate as a function of the shadow radius. Its behavior in $(\omega_{max} - r_s)$ plane is exactly similar to that already displayed in  Fig.~\ref{rsrh} in $(T - r_h)$ plane: For  $P>P_c$ the maximum frequency is just a monotone increasing function of $r_s$, with no extremum.  As for $P<P_c$, The isobaric frequency plot has both a local maximum and local minimum, which imply a non monotonic behavior indicating a first order small-large black hole phase transition. When the pressure approaches $P_c$, these two local extremums merge into an inflection point.

\begin{figure}[h]
  \begin{center}
    \includegraphics[width=.48\linewidth]{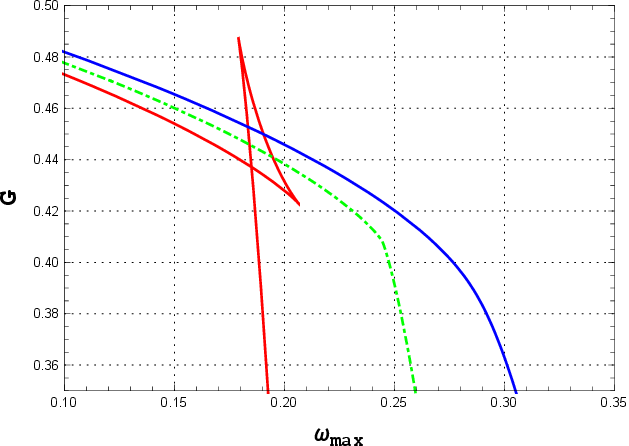}
    \includegraphics[width=.48\linewidth]{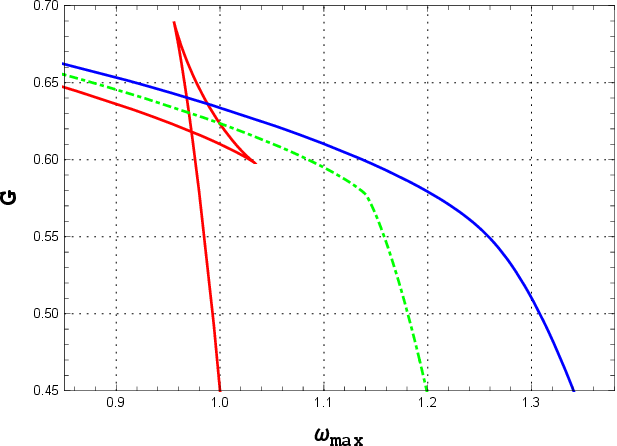}
       \caption{Gibbs free energy  as a function of  the maximum emission rate for an  isobaric process. \textbf{Left:} RN-AdS black hole without massive gravity. \textbf{Right:} RN-AdS black hole in massive gravity with $a=1$ and $b=1$. The following inputs are used  $q=1$ and $r_0=20$.}
   \label{Gwm}
  \end{center}
\end{figure}

Moreover, we also analyze in  Fig.~\ref{Gwm} the Gibbs free energy $(G=H-TS)$ as a function of  the maximum frequencies.  We observe a swallowtail behavior that generally characterize Van der Waals system.  Again, we get a phase structure similar to that shown by Fig.~\ref{GT} in $G-T$ plane. This consolidate further our new prescription, trading  $\omega_{max}$ with the black hole temperature,  as a reliable and powerful means to deal with thermodynamics and phase transitions of RN-AdS black holes in massive gravity. 

At this stage, note that the result reported in \cite{Kubiznak:2012wp,Hamil:2023dmx} is reproduced for $a=0$ and $b=0$, as well as the famous relation $P_c\, v_c/T_c=3/8$.
 
\begin{figure}[h]
  \begin{center}
    \includegraphics[width=.47\linewidth]{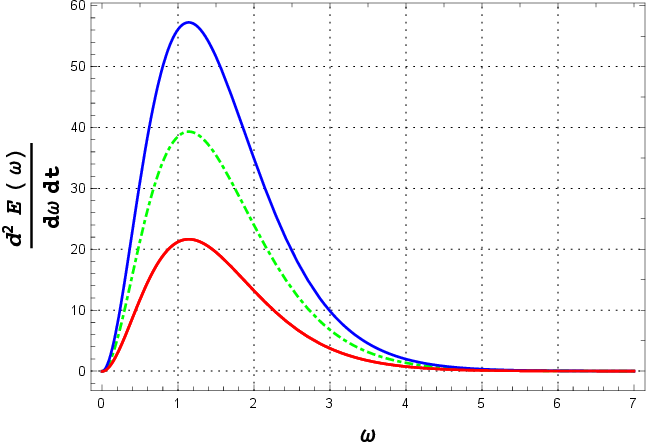}
    \includegraphics[width=.5\linewidth]{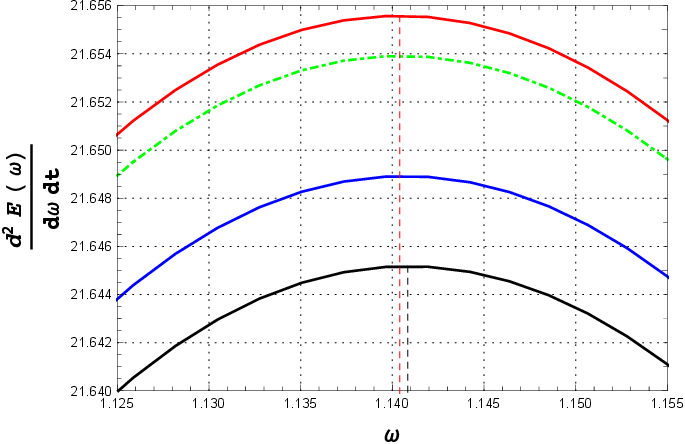}
       \caption{\textbf{Left}: The emission rate of massive field as a function of the frequency for black hole subcritical phase (red), supercritical phase (blue) and critical point (green dashed). The temperature and mass of emitted particles are fixed. \textbf{Right:} The  maximum frequency for different  particle masses: red for $m=0$, green for $m=0.01$, blue  for $m=0.02$  and black  for $m=0.03$. We set $a = 1$, $b=1$, $q=1$, $r_0=20$ with $M=1$. }
   \label{Emm}
  \end{center}
\end{figure}

 In the subsequent part of this work, we address the case where the  emitted   fields are massive. In this context, the emission rate per unit frequency in four dimensions is given by the formula\footnote{Note that this formula is only valid for bosonic modes, The minus sign in the denominator must be modified for  fermionic modes.}~\cite{Emparan}.

 \begin{equation}
 \frac{d^2E(\omega,m)}{d\omega dt}\simeq \left(\omega^2 -m^2\right)\omega\frac{\pi r_h^2}{e^{\omega/T_H} -1}.
 \label{emmassive}
 \end{equation}
  Knowing that a black hole acts as a perfect absorber of massive particles,  the horizon area is used as a good approximation of the absorption cross section. Fig.~\ref{Emm} displays  the energy emission rate as a function of  frequency.  The same features  as for the massless emitted particle are observed: existence of a peaks where the maximum frequencies increase when the particle mass gets larger. However, we see instead a decreasing of the rate for these frequencies,  which could be useful for observation, as the low-mass particles will be more likely to be detected, especially if  quantum black holes in highest-energy particle colliders are detected.

      Unlike the massless emitted particle (photons)  where  the classical Wein's law is exactly reproduced, solution of Eq.~\ref{emmassive} for massive  emitted particles introduces corrections which yield a generalized Wein's displacement law :
    \begin{equation}
    \omega_{max}(m,T_H)\approx 0.60\, T_H + \frac{0.28\, m^2 + 1.11\, T_H^2}{\left(0.37\, m^3 + 0.94\, T_H^3\right)^{1/3}} + 1,11\left(0.37\, m^3 + 0.94\, T_H^3\right)^{1/3}.
        \label{wcor}
    \end{equation}
 If $m=0$,  one can readily reproduce the classical Wein's law,   $\omega_{max}(m \approx 0,T_H)\approx 2.82\, T_H$. 
 
 \begin{figure}[h]
  \begin{center}
    \includegraphics[width=.47\linewidth]{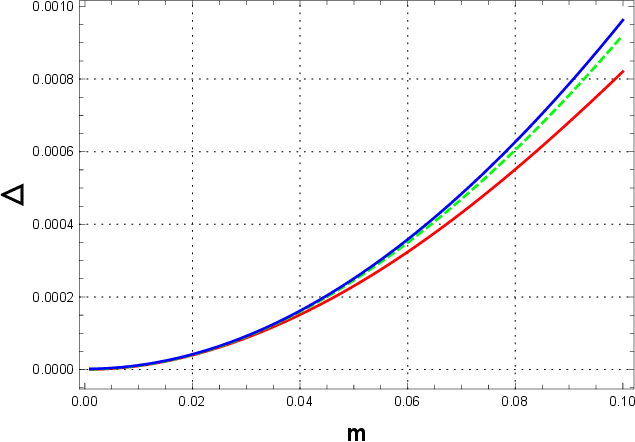}
    \includegraphics[width=.47\linewidth]{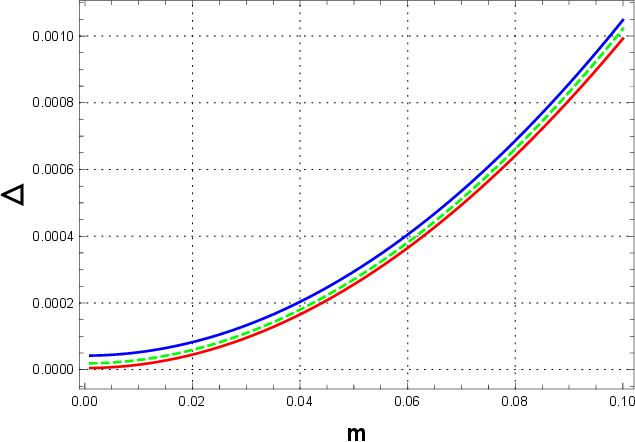}
       \caption{Relative deviation from the massless fields versus emitted particles masses $m$  in the  isotherm transformation: red line for $T=0.5\,T_H$, green for $T = T_H$, and blue  for $T=1.5\,T_H$.  We set $q=1$. \textbf{Left}: $a=0$ and $b=0$. \textbf{Right}: $a=1$ and $b=1$.}
   \label{diff}
  \end{center}
\end{figure}

In order to  perceive to what extent our result for massive emitted particles affects the Wien's law with respect to the massless case (photons), we analyze the relative deviation expressed by the equation, 
\begin{equation}
\Delta=\frac{|\omega_{max}(m)-\omega_{max}(0)|}{\omega_{max}(0)}.
\end{equation}

Fig.~\ref{diff} shows $ \Delta$ with respect to the particle mass $m$. For small masses, below the Hawking temperature, no noticeable effect is observed, so the Wein's displacement law is almost insensitive to the particle mass. In contrast, when the mass tends to or gets larger than $T_H$ the relative deviation is significant  increasing linearly with the emitted particle mass. This behavior persists in presence of massive gravity or without.

\begin{figure}[h]
	\begin{center}
		\includegraphics[width=.47\linewidth]{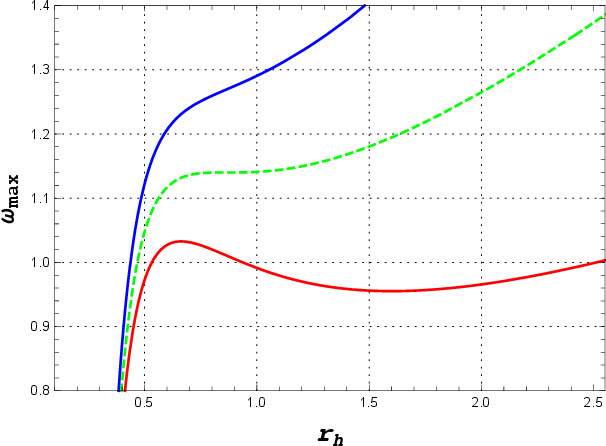}
		\includegraphics[width=.47\linewidth]{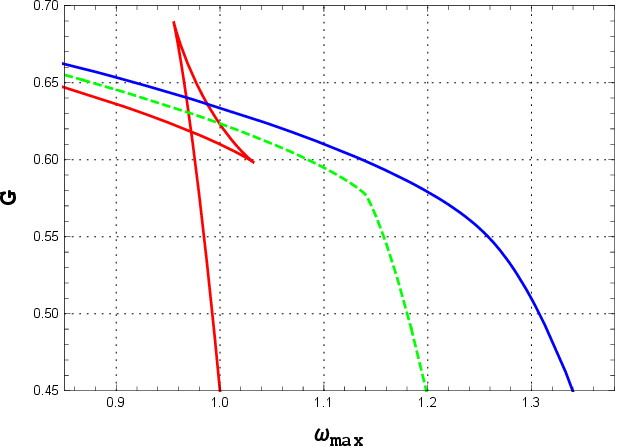}
		\caption{\textbf{Left}: $\omega_{max}(m)-r_h$ diagrams in isobaric process. \textbf{Right}: $G -\omega_{max}(m)$ diagrams in isobaric process.	The red line for $P=0.5\,P_C$, green represents $P = P_C$, blue line is for $P=1.5\,P_C$.  We set $q=1$, $m=0.01$, $a=1$ and $b=1$. }
		\label{wmcor}
	\end{center}
\end{figure}

Lastly, we plot  in Fig.~\ref{wmcor} the maximum frequency of massive field as a function of the horizon radius and the Gibbs free energy versus the maximum emission frequency. As expected a thermodynamic behavior similar to that obtained for photons is seen. So this analysis might open a new window towards detecting thermodynamic phase transition using the emission spectrum of massive particles, especially if  a micro black holes  are observed in the high energy particle colliders.

 \FloatBarrier
\section{Conclusion}
\label{Conclusion}

In this paper, we have proposed an approach relying on the black hole shadow as a relevant tool to uncover the thermodynamics phase transition in massive gravity. Besides being a physical observable that can be probed by means of  observation such as  M87* image~\cite{EHT:2019nmr}, the shadow radius $r_s$ is capable to replicate the horizon radius $r_h$ and entropy in describing black holes criticality. Indeed, since  $r_h$ and  $r_s$ are perfectly correlated,  the analysis of black hole thermodynamics turn out to be exactly similar whether it is performed in $T-r_h$, $T-S$ or  $T-r_s$ planes., though  the last one with the shadow radius comes with additional remarkable feature related to observation.\\

Moreover, we have investigated the energy emission rate for electromagnetic spectrum emitted within the shadow region of  the black hole in massive gravity. Emphasis were placed in the frequency $\omega_{max}$ corresponding to the maximum energy emission rate from which we developed a prescription to examine AdS black holes thermodynamics in massive gravity. Our analysis showed that by trading the usual quantities $T$ and $r_h$ with the observable ones $\omega_{max}$ and $r_s$,  we accurately reproduce all the thermodynamics and phase transition features of  black hole. \\

Finally, since most of the energy is radiated on the brane where the emitted mode is either a photon or a massive bosonic field, we also  studied the energy emission rate of massive particles for RN-AdS black hole within massive gravity. We found that the radiation spectrum may affect the thermodynamical criticality and induce deviation from the Wein's law when the particle mass is larger than Hawking temperature. \\

{{\bf Data Availability Statement}: No Data associated in the manuscript. 
}

\end{document}